\documentclass[10pt,letter]{article}

\usepackage{amsmath}
\usepackage{comment}

\usepackage{caption}
\usepackage{fixltx2e}
\usepackage{float}
\usepackage[margin=2.5 cm]{geometry}
\usepackage{graphicx}
\usepackage[utf8]{inputenc}
\usepackage{setspace}
\usepackage{textgreek}
\usepackage{titlesec}
\usepackage{gensymb}
\usepackage{enumitem}

\makeatletter
\renewcommand\@biblabel[1]{}
\makeatother

\renewenvironment{abstract}
 {\par\noindent\textbf{\abstractname}\ \ignorespaces}
 {\par\medskip}

\titleformat{\section}
  {\normalfont\fontsize{10}{15}\bfseries}{\thesection.}{1em}{}
  
\titleformat{\subsection}
  {\normalfont\fontsize{10}{15}}{\thesubsection.}{1em}{}
\titleformat{\subsubsection}
            {\normalfont\fontsize{10}{15}}{\thesubsubsection.}{1em}{}

\begin{document}
\begin{flushleft}
\normalsize 
\textbf{\large A Python Code to Determine Orbital Parameters of Spectroscopic Binaries}\\[5ex]
%
%
\textbf{\normalsize Nicholas Milson, Caroline Barton, and Philip D.\ Bennett\\[1ex]}
{\em Department of Physics and Atmospheric Science, Dalhousie University, Halifax, NS~ B3H 4R2, Canada}\\
\end{flushleft}

\vspace{1ex}

\pagenumbering{arabic}

\begin{abstract}
We present the open source Python code {\em BinaryStarSolver} that solves for the orbital elements of a spectroscopic binary system. Given a time-series of radial velocity measurements, six orbital parameters are determined: the long-term mean, or systemic, radial velocity, the velocity amplitude, the argument of periastron, the eccentricity, the epoch of periastron, and the orbital period referred to by $\{{\gamma, K, \omega, e, T_0, P}\}$ respectively. Also returned to the user is the projected length of the semi-major axis, $a_{1}\sin(i)$, and the mass function, $f(M)$. The determination of spectroscopic orbits and masses is an example of another important  area of astrophysics, once the domain of professional astronomers, to which amateurs can now make significant contributions. This code, available from GitHub, is provided in support of that work, and should be of general use to the amateur and professional astronomical community.
\end{abstract}

\section{Introduction}

With the advent of commercially available research-grade spectrographs, the amateur astronomy community, especially in Europe, is now acquiring professional quality spectroscopic and radial velocity data for variable and binary stars, see, e.g, Pollmann \& Bennett (2020), and this trend is likely to grow over the coming years. The determination of spectroscopic orbits and masses is an example of another important  area of astrophysics, once the domain of professional astronomers, to which amateurs can now make significant contributions. Here we present a summary of the problem of determining orbital parameters from a time series of radial velocity observations of a spectroscopic binary star, and provide the astronomical community with a new open source Python code, {\em BinaryStarSolver}, to carry out this orbit solution.\\

The orbit of a star in a binary system is uniquely specified by seven orbital parameters. These are the eccentricity $e$, the semi-major axis $a$, the time (or epoch) of periastron $T_0$, the orbital period $P$, the argument of periastron $\omega$, the position angle of the ascending node $\Omega$, and the inclination angle $i$. For a spectroscopic binary system, radial velocities of the stars can be inferred from analysis of their Doppler-shifted spectral lines. The radial velocity, $V(t)$, of the primary star in a binary system as a function of time $t$ is given by (Green 1985),

\begin{equation}
  V(t) = \gamma + \frac{na_1\sin i}{\sqrt{1-e^2}} \,\, \big\{\cos[v(t)+\omega] + e\cos\omega\big\}
\label{V1}
\end{equation}

where $n=2\pi/P$ is the mean motion, $a_1$ is the semi-major axis of the primary star's orbit, $\gamma$ is the long term mean or systemic velocity of the binary,  and $v(t)$ is the true anomaly as a function of time. It can be seen in this expression that $a_1$ and $i$ are coupled. A separate determination of each parameter is not possible without additional information, independent of radial velocities. Therefore, it is standard practice to define the new parameter, the radial velocity semi-amplitude of the primary star $K_1$, as

\begin{equation}
  K_1 = \frac{na_1 \sin i }{\sqrt{1-e^2}}.
\label{Kdef}
\end{equation}

The velocity semi-amplitude of the secondary star in the binary, $K_2$, is defined in an analogous manner using $a_2$. The expression for radial velocity as a function of time for the $j$-th star ($j=1,2$) in the binary system then becomes 

\begin{equation}
    V_j(t) = \gamma + K_j\big\{\cos[v(t)+\omega] + e\cos\omega \big\}, \hspace{0.10in} j=1,2
\label{V2}
\end{equation}

If the spectra of both stars can be observed and separated, then we have the case of an SB2 (double-lined) spectoscopic binary. In this case, knowledge of $K_1$ and $K_2$ of both stars suffices to determine the ratio of stellar masses: $M_1/M_2 = K_2/K_1= a_2/a_1$. Without additional information about the inclination angle $i$, individual stellar masses can not be found; the best that can be accomplished is the determination of the mass function 
$f(M) =  (M_1 + M_2) \sin^3 i$. \\

An additional constraint is provided by Kepler's third law (Green 1985). 
\begin{equation}
  M_1 + M_2 = \frac{4\pi^2 a^3}{G P^2} = 3.985 \times 10^{-20} \, \frac{(a_1 + a_2)^3}{P^2}
\label{Kepler}
\end{equation}
where $a = a_1 + a_2$ is the semi-major axis of the relative orbit of the two stars, and the constants have been evaluated to yield masses in solar units ($M_\odot$) given $a_j$ in km and $P$ in days. Then, the mass function can be written
\begin{equation}
f(M) = (M_1 + M_2) \sin^3 i = 3.985 \times 10^{-20} \, \frac{(a_1\sin i + a_2\sin i)^3}{P^2}
\label{Kepler2}
\end{equation}
where 
\begin{equation}
a_j \sin i = \frac{K_j \sqrt{1-e^2}}{n}, \hspace{0.10in} j=1,2
\label{asini}
\end{equation}\\

For the special case of eclipsing binaries, the inclination $i$ must be close to $90^\circ$ for eclipses to occur, and in that case we can typically assume $\sin i = 1$. Then, the total mass $M_1 + M_2$, and the mass ratio $M_1/M_2$ are both known, and the individual stellar masses $M_1$ and $M_2$ are determined, as are $a_1$ and $a_2$. Therefore, a complete determination of the orbit and masses of the stars in an eclipsing, double-lined spectroscopic binary (SB2) is possible.\\

For convenience, we will from now on normally use $K$ without a subscript to refer to either star, and if only one star is implied, the context will make this clear. Notice that time only appears in this expression implicitly. Since the measurements made are a function of time, not true anomaly, a relationship between the two variables is needed. This follows from two equations: first, an expression between the true anomaly and the eccentric anomaly, $E$, and second, between the eccentric anomaly and time (the latter is known as Kepler's Equation, Green 1985).

\begin{equation}
    \tan \frac{v}{2} = \frac{1+e}{1-e} \tan \frac{E}{2}
\label{E1}
\end{equation}
\begin{equation}
    t = \frac{E-e\sin E}{n} \, + \, T_0
\label{t1}
\end{equation}\\

 Note that $\Omega$ does not appear in any of the previous expressions. This is to be expected since $\Omega$ specifies the orientation of the orbit in the plane of the sky, and so has no effect of the observed radial velocity. This also means that it cannot be determined solely from radial velocity measurements. Therefore, we do not concern ourselves further with $\Omega$ here. This leaves six parameters to be determined: $\{\gamma, K, \omega, e, T_0, P\}$. \\
 
 In the case of an eclipsing binary, the period $P$ may be found more accurately through the observation of eclipses. If this is the case, in the provided code, the user can opt to provide the known period and solve for only $\{\gamma,K,\omega,e,T_0\}$. Generally though, $P$ is not predetermined and the following discussion assumes this case.

\section{Discussion of procedure}
\subsection{Minimization}
To determine the orbital parameters, the given radial velocity data is fit with a curve of the form of (\ref{V2}), using a nonlinear least-squares approach to minimize the residual:
\begin{equation}
\chi^2 = \sum_{i=1}^{N} w_i \bigg(V_i - \Big\{\gamma + K\big[\cos(v(t_i)+\omega) + e\cos\omega\big]\Big\}\bigg)^2,
\label{Res}
\end{equation}

where $V_i$ is the observed radial velocity at time $t_i$, and $w_{i}$ is the weight of the measurement.\\

To carry out this minimization, the Levenberg-Marquardt algorithm (also known as the damped least-squares method) is used. This algorithm combines Newton's method and the gradient descent method through the introduction of a dampening parameter $\lambda$ to give both a more efficient and reliable convergence than either method on their own (Gavin 2011, Marquardt 1963).\\

Specifically, in this method, the entries along the diagonal of the Hessian matrix, $\boldsymbol H$, are multiplied by $(1 + \lambda)$. Refer to Appendix A for the full Hessian matrix. When the parameter $\lambda$ is large ($\lambda \gg 1$), then the Hessian matrix ${\boldsymbol H} \approx \lambda {\boldsymbol I}$, where $\boldsymbol I$ is the $6 \times 6$ diagonal identity matrix, and then the change in parameter vector $\Delta \boldsymbol p$ follows the gradient descent direction. As $\lambda \rightarrow 0$, the change $\Delta \boldsymbol p$ approaches that of Newton's method. Since a change in the direction of the negative of the gradient ensures a reduction in the local value of $\chi^2$, an initial estimate should be made using a suitably large value of $\lambda$.\\

Apart from the introduction of the $\lambda$ coefficient, however, the minimization procedure closely follows Newton's method. So, where the current estimate of parameters is the vector, $\boldsymbol p_n$, $\boldsymbol p_{n+1}$ may be found by solving for $\Delta \boldsymbol p = \boldsymbol p_{n+1} - \boldsymbol p_n$ from the system of equations

\begin{equation}
    \big[\boldsymbol H\chi^2(\boldsymbol p_n)\big]\Delta \boldsymbol p =  - \nabla \chi^2(\boldsymbol p_n) ,
    \label{lv}
\end{equation}
where $\nabla \chi^2(\boldsymbol p_n)$ is the gradient of $\chi^2$ at $\boldsymbol p_n$ (Press et al. 1992). Complete details of the Hessian matrix and the partial derivatives used in the minimization are given in Appendix A.\\

Upon running the minimization procedure, an initial of $\lambda = 3$ is used. This initial value of $\lambda$ was found to be a reasonable compromise between ensuring convergence while retaining accuracy of the final parameter values. If after one iteration the residual decreases, $\lambda$ is reduced by a factor of 9. If instead the residual increases, $\lambda$ is increased by a factor of 11. In this case, the current parameter vector $\boldsymbol p_{n+1}$ is rejected and the next iteration starts using the previous parameter vector $\boldsymbol p_n$ again (Gavin 2011, Press et al. 1992, Transtrum \& Sethna 2012).
The minimization procedure is then left to run until the change in the residual becomes negligible (either an absolute change $<0.01$, or a fractional change $<10^{-3}$ ), at which point sufficient convergence is achieved (Press et al. 1992).\\

The parameters returned are then used to calculate $a_1\sin(i)$ and the mass function (see Appendix A) and the stellar parameters $\{\gamma, K, \omega, e, T_0, P, a_1\sin i, f(M)\}$ are returned to the user with their respective uncertainties, as discussed in Section 2.4.\\

\subsection{Companion star}
Included in this code, as well, is a separate function that solves for the orbital parameters of a star given the parameters of its companion (and radial velocity data). This allows for a more accurate determination of the orbital parameters of the star if less data is available for it compared to its companion.\\

In a binary star system, the orbital elements of the companion star are nearly all identical to that of the primary star. The eccentricity, inclination angle, orbital period, and epoch of periastron passage are all the same for both stars and the argument of periastron of the companion star only differs from the argument of periastron for the primary star by 180\degree (Green 1985). This leaves the semi-major axis, $a_2$, (and by extension the velocity amplitude $K_2$) as the only undetermined parameter. Since $K_2$ is the only variable parameter, it is determined using Newton's method.\\

As arguments, the companion function takes the determined orbital elements $\{{\gamma, K_1, \omega_1, e, T_0, P}\}$ of the primary star, and returns the orbital elements $\{\gamma, K_2, \omega_2, e, T_0, P, a_2\sin i, f(M)\}$ of the companion star, requiring only a few radial velocity points and their respective times to accurately determine $K_2$.\\

\subsection{Initial estimates}
Like other minimization methods, the Levenberg-Marquardt algorithm will often fail to find the global minimum if the initial estimates are not accurate enough. Thus, unless initial estimates are provided by the user, the minimization routine begins by carefully approximating the parameters $\{\gamma, K, \omega, e, P, T_0\}$ from the given data.\\

First, $\gamma$ can be estimated using the average of the radial velocity data, $V_{avg}=\sum V_i/N$, where $N$ is the number of data points. To approximate the other parameters, equations (\ref{V2}), (\ref{E1}), and (\ref{t1}) must be manipulated. Beginning with equation (\ref{V2}), it can be seen that the maximum radial velocity occurs when $\cos(v + w)=1$ and the minimum radial velocity occurs when $\cos(v + w)=-1$ (Green 1985). At $V_{max}$ and $V_{min}$, (\ref{V2}) then reduces to two equations

\begin{equation}
    V_{max} = \gamma + K\big[1 + e\cos w\big]
    \label{Vmax}
\end{equation}
\begin{equation}
    V_{min} = \gamma + K\big[-1 + e\cos w\big]
    \label{Vmin}
\end{equation}

By adding/subtracting equations (\ref{Vmax}) and (\ref{Vmin}), it can be shown that

\begin{equation}
    \frac{V_{max} -V_{min}}{2} = K
    \label{K}
\end{equation}
\begin{equation}
    \frac{V_{max} +V_{min}}{2} = \gamma + Ke\cos w
    \label{w}
\end{equation}

From equation (\ref{K}), $K$ can be estimated using $V_{max}$ and $V_{min}$ from the data.

$T_0$ and $\omega$ can be related to each other through two equations. First by rearranging equation (\ref{E1}),

\begin{equation}
    E_{max}= 2\arctan\bigg[{\Big(\frac{1+e}{1-e}\Big)}^{-\frac{1}{2}}\tan\Big(\frac{v_{max}}{2}\Big)\bigg]
    \label{Emax}
\end{equation}
Where $E_{max}$ and $v_{max}$ are the eccentric and true anomaly when $V = V_{max}$ (which may be found from equations (\ref{E1}) and (\ref{t1})). However, $\cos(v + \omega) = 1$ at $V_{max}$, $v_{max} = -\omega$ and so,
\begin{equation}
    E_{max} = 2\arctan\bigg[{\Big(\frac{1+e}{1-e}\Big)}^{-\frac{1}{2}} \tan\Big(\frac{-\omega}{2}\Big)\bigg]
    \label{t2}
\end{equation}

Next, rearranging equation (\ref{t1}), we find

\begin{equation}
    T_0 = t_{max} - \frac{1}{n}\big[E_{max} - e\sin E_{max}\big]
\end{equation}
Here $t_{max}$ is the time corresponding to $V_{max}$.

Now, armed with these equations, initial estimates may be found using the method outlined in the following list.
\begin{enumerate}[noitemsep]
    \item Estimate $\gamma$ by averaging the radial velocity data
    \item Estimate $K$ using (\ref{K})
    \item Unless $P$ is supplied by the user, find $P$ using two methods (further described in 2.3.1)
    \begin{enumerate}
        \item From the two points lying close to the same $V$ axis, separated by an integer number of periods
        \item By averaging the separation between maxima, minima, and points crossing the same $V$ axis
    \end{enumerate}
    \item Iterate $e$ from 0.01 to 0.99 with a step size of 0.01 and for each value
    \begin{enumerate}
        \item Find $\omega$ using (\ref{w})
        \item Find $T_0$ using (\ref{Emax}) and (\ref{t2})
        \item Using the estimates for $\{\gamma, K, \omega, e, T_0\}$ and
        \begin{enumerate}
            \item $P$ from 3(a), calculate the sum of the squared deviations
            \item $P$ from 3(b), calculate the sum of the squared deviations
        \end{enumerate}
        \item If either of 4(c)i.\ or 4(c)ii.\ result in a smaller residual than the previous best estimate of the parameters, take the $\{\gamma, K, \omega, e, T_0, P\}$ used as the new best estimate
    \end{enumerate}
\end{enumerate}

\subsubsection{Estimating the period}
If $P$ is not provided by the user, it must also be estimated. Both estimates of the period mentioned in 3(a) and 3(b) begin by finding each point where the data crosses a given line. The line used here is the $V_{avg}$ axis, also the estimate for $\gamma$. A point $V_i$ is deemed a `crossing point' if $V_{i-1}$ and $V_{i+1}$ lie on opposite sides of the axis formed by $V_{avg}$. Thus, unless the data is messy where it crosses the axis or $V_i = V_{avg}$, crossing points will occur in pairs. A common error caused by messy data is that the data crosses $V_{avg}$ twice but not in a sequential pair, as demonstrated in Figure 1.
It is assumed that there will be sufficient data that if the crossing points are on opposite sides of an extremum in the true radial velocity curve, at least one point in between will exceed a displacement of $\frac{\text{1}}{\text{6}}^{\text{th}}$ the amplitude of the data, $\frac{V_{max} - V_{min}}{2}$, from $V_{avg}$. So, for all crossing points that do not occur sequentially in the data, the code verifies that there is at least one data point in between that meets the above criteria. If this is not the case, whichever of the two crossing points lay further from the $V_{avg}$ axis is discarded and replaced by the data point directly before (if the first point is discarded) or after (if the second point is discarded) the point that was kept.\\

After finding the crossing points, the first method estimates the period by finding two crossing points, separated by an integer number of periods, that have the closest radial velocity values. The period is then found from these time values and the integer number of periods between the two points.\\

In the second method, once the crossing points are found, if the points span over one period the extremum between every pair of points is found. Now $P$ may be found by averaging all estimates of the period from the crossing data as well as the extrema data.\\

Now both estimates for $P$ may be used in the iterations of $e$.\\

\begin{figure}[H]
    \centering
    \includegraphics[scale=0.9]{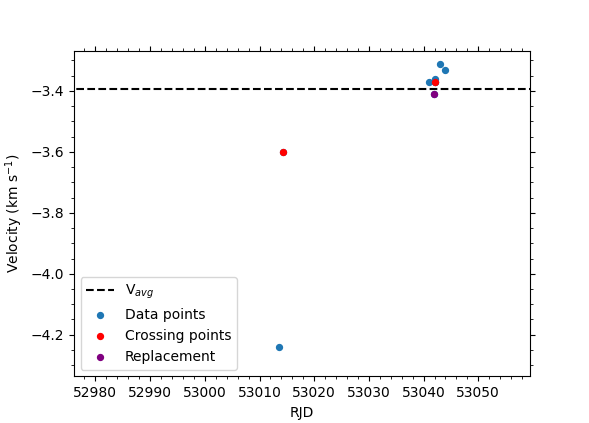}
    \caption{Example of two non-consecutive points that cross the $V_{avg}$ axis. Here the second red point would be kept and the first would be replaced by the purple point.}
\end{figure}

\subsection{Error analysis}
Once the parameters $\boldsymbol{p}$ have been determined, the uncertainties for each of the six (or five if $P$ is known) fitted parameters are computed and returned as well.

First, the covariance matrix, $\boldsymbol{C}$, is given by inverting the Hessian matrix, i.e.

\begin{equation}
    \boldsymbol{C} = \boldsymbol{H}^{-1}
\end{equation}

The diagonal elements of the covariance matrix, $\boldsymbol C_{kk}$, give the variance of parameters $\boldsymbol p_k$. Thus, the square root of the diagonal elements gives the standard error in the associated parameters (Press et al. 1992). I.e.,

\begin{equation}
    \sqrt{\boldsymbol C_{kk}} = \sigma_{k}
\end{equation}

The off diagonal elements, $\boldsymbol C_{mk}$, where $m \neq k$, gives the covariance of parameters $\boldsymbol p_k$ and $\boldsymbol p_m$. The linear coefficient of correlation for two given parameters is defined as 
\begin{equation}
    r_{mk} = \frac{\sigma_{mk}}{\sigma_{m}\sigma_{k}}
\end{equation}

Therefore, the correlation coefficient $r_{mk}$ can be determined from the covariance matrix by
\begin{equation}
    r_{mk}= \frac{\boldsymbol C_{mk}}{\sqrt{\boldsymbol C_{mm}}\sqrt{\boldsymbol C_{kk}}}
\end{equation}
(Press et al. 1992).

\section{Comparison of results}
Here the {\em BinaryStarSolver} code is used to determine the orbital elements of 31 Cygni, 32 Cygni, and Tau Persei using R. F. Griffin's 2008 and R. E. M. Griffin's 1992 data, with the same weights as originally published. The results are compared to the original solutions to show the effectiveness of this code.\\

\begin{table}[H]
    \centering
	\caption{Orbital Elements of 31 Cygni, from Griffin (2008)}
\begin{tabular}{ c c c  }
    \hline
    Orbital Element & Griffin Solution & Solution, this paper \\
    \hline
    $P$ (days) & $3784.3$  &$3784.3$  \\
    $\gamma$ (km/s)& $-6.421 \pm 0.034$   &$-6.425 \pm 0.032$  \\
    $K$ (km/s) & $ 13.94 \pm 0.04$ &$ 13.94 \pm 0.04$   \\
    $\omega$ (\degree) & $204.5 \pm 1.0$  & $204.5 \pm 1.0$ \\
    $e$  & $0.2084 \pm 0.0031$  & $0.2080 \pm 0.0029$  \\
    $T_0$ (RJD)&  $52346 \pm 9$ & $52346 \pm 9$   \\
    $a\sin{i}$ (Gm) & $709.5 \pm 2.3$ & $709.3 \pm 2.1$  \\
    $f(m)$  $(\textup{M}_\odot)$ & $0.995 \pm 0.010$ & $0.993 \pm 0.009$ \\ 
    \hline
\end{tabular} \\
\smallskip
 \small{Note -- Here, the period is kept fixed, as it is in Griffin (2008).}
 \end{table}

\begin{figure}[H]
    \centering
    \includegraphics[scale=0.9]{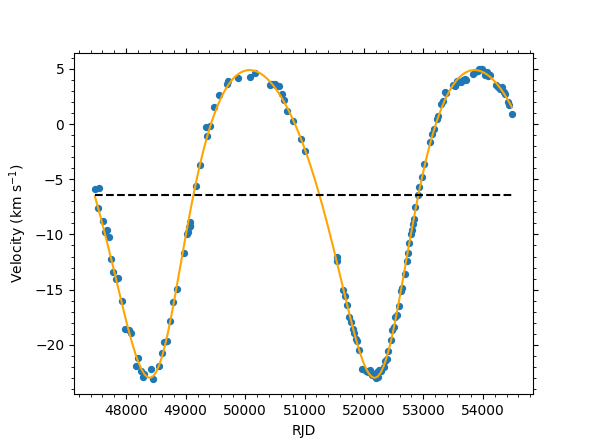}
    \caption{31 Cygni radial velocity observations of Griffin (2008) plotted as a function of time, with the velocity curve using the parameters found by this code drawn through the points.}
\end{figure}

\begin{table}[H]
    \centering
    \caption{Orbital Elements of 32 Cygni, from Griffin (2008)}

\begin{tabular}{ c c c  }
    \hline
    Orbital Element & Griffin Solution & Solution, this paper \\
    \hline
    $P$ (days)  & $1147.51 \pm 0.31 $  & $1147.58 \pm 0.31$   \\
    $\gamma$ (km/s) & $-6.389 \pm 0.032$& $-6.393 \pm 0.031$    \\
    $K$ (km/s) & $ 16.77 \pm 0.05$  & $16.76 \pm 0.05$  \\
    $\omega$ (\degree)  & $221.4 \pm 0.5 $   & $221.5 \pm 0.5$  \\
    $e$  & $0.3041 \pm 0.0027$  & $0.3040 \pm 0.0027$   \\
    $T_0$ (RJD) &  $52647.4 \pm 1.4$ &   $52647.6 \pm 1.4$   \\
    $a\sin{i}$ (Gm) & $252.0 \pm 0.8$  & $252.0 \pm 0.8$   \\
    $f(m)$  $(\textup{M}_\odot)$ & $0.486 \pm 0.005$  & $0.484 \pm 0.005$ \\ 
    \hline
\end{tabular}
\end{table}

\begin{figure}[H]
    \centering
    \includegraphics[scale=0.9]{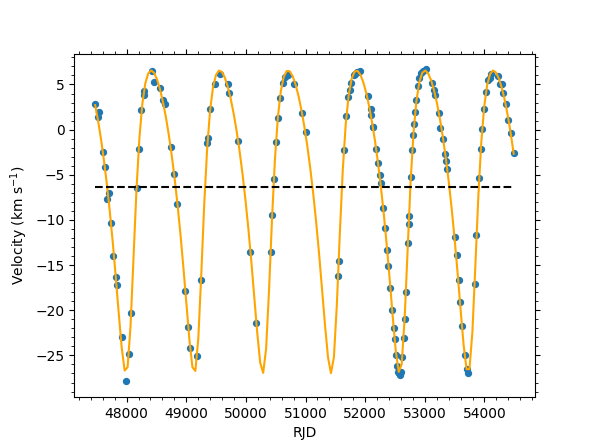}
    \caption{32 Cygni radial velocity observations of Griffin (2008) plotted as a function of time, with the velocity curve using the parameters found by this code drawn through the points.}
\end{figure}

\begin{table}[H]
    \centering
    \caption{Orbital Elements of $\tau$ Persei, from Griffin et al.\ (1992)}
\begin{tabular}{ c c c  }
    \hline
    Orbital Element & Griffin et al.\ Solution & Solution, this paper \\
    \hline
    $P$ (days)  &  $1516.1 \pm 1.9$  & $1517.0 \pm 2.6$    \\
    $\gamma$ (km/s) & $2.37 \pm 0.11$  & $2.34 \pm 0.15$  \\
    $K$ (km/s) & $18.7 \pm 0.3$ & $18.7 \pm 0.4$  \\
    $\omega$ (\degree)  &  $234.7 \pm 1.3$  &  $233.7 \pm 1.5$ \\
    $e$  & $0.721 \pm 0.007$ & $0.719 \pm 0.009$  \\
    $T_0$ (RJD) & $47524.1 \pm 1.4$  & $46006.7 \pm 3.4$   \\
    $a\sin{i}$ (Gm) & $271 \pm 5$ & $270 \pm 7$  \\
    $f(m)$  $(\textup{M}_\odot)$& $0.344 \pm 0.020$  & $0.342 \pm 0.025$  \\ 
    \hline
\end{tabular}%

\end{table}

\begin{figure}[H]
    \centering
    \includegraphics[scale=0.9]{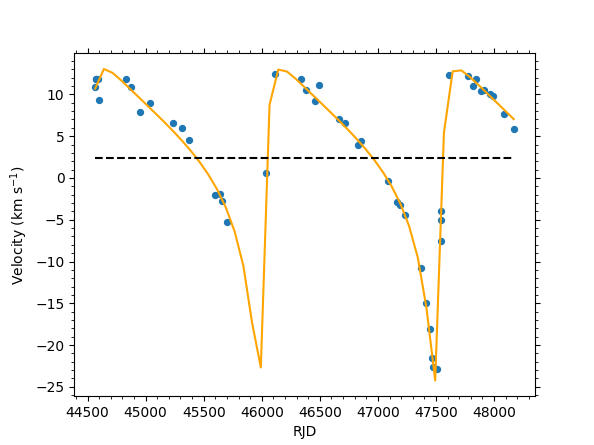}
    \caption{$\tau$ Persei radial velocity observations of Griffin et al.\ (1992) plotted as a function of time, with the velocity curve using the parameters found by this code drawn through the points. This star, with its highly eccentric ($e=0.72$) orbit, provides a challenging test of convergence for the {\em BinaryStarSolver} code.}
\end{figure}

\section{Conclusions}
In this paper, we present a new, open source Python code {\em BinaryStarSolver} to determine spectroscopic binary orbits, that is readily accesible to both amateur and professional astronomers. The determination of spectroscopic binary orbits and stellar masses is an example of another important area of astrophysics to which amateurs can now make significant contributions, but analysis software accessible to the amateur community remains a constraint. The present code is intended to provide a robust, easy-to-use package to determine binary star orbits given a time series of radial velocity observations.\\

We describe the solution procedure used by the code to determine binary star orbital elements, and present results for three binary systems with recent analyses in the literature. One of these binaries ($\tau$~Per) has a highly eccentric orbit ($e=0.72$), which renders the problem less stable numerically. Our orbital solutions for all three binaries are in extremely close agreement with the previously published orbits, demonstrating the accuracy and robustness of the method used by the {\em BinaryStarSolver} code.\\


The {\em BinaryStarSolver} code described in this paper is available, with complete documentation, from GitHub\footnote{\tt https://github.com/NickMilsonPhysics/BinaryStarSolver}. We also plan to have the code included on the American Association of Variable Stars Observers (AAVSO) software directory page.

\appendix
\section{Appendix: Summary of Equations}

The Hessian matrix\footnote{If $P$ is known, the Hessian reduces to a 5x5 matrix where the 6th row and 6th column are deleted.}:
\[\boldsymbol{H} =
\begin{bmatrix}
     \frac{\partial ^2 \chi^2}{\partial \gamma^2} & \frac{\partial ^2 \chi^2}{\partial \gamma \partial K} & \frac{\partial ^2 \chi^2}{\partial \gamma \partial \omega} & \frac{\partial ^2 \chi^2}{\partial \gamma \partial e} & \frac{\partial ^2 \chi^2}{\partial \gamma \partial T_0} & \frac{\partial ^2 \chi^2}{\partial \gamma \partial P} \\
     \frac{\partial ^2 \chi^2}{\partial K \partial \gamma} & \frac{\partial ^2 \chi^2}{\partial K^2} & \frac{\partial ^2 \chi^2}{\partial K \partial \omega} & \frac{\partial ^2 \chi^2}{\partial K \partial e} & \frac{\partial ^2 \chi^2}{\partial K \partial T_0} & \frac{\partial ^2 \chi^2}{\partial K \partial P} \\
    \frac{\partial ^2 \chi^2}{\partial \omega \partial \gamma} & \frac{\partial ^2 \chi^2}{\partial \omega \partial K} & \frac{\partial ^2 \chi^2}{\partial \omega^2} & \frac{\partial ^2 \chi^2}{\partial \omega \partial e} & \frac{\partial ^2 \chi^2}{\partial \omega \partial T_0} & \frac{\partial ^2 \chi^2}{\partial \omega \partial P} \\
    \frac{\partial ^2 \chi^2}{\partial e \partial \gamma} & \frac{\partial ^2 \chi^2}{\partial e \partial K} & \frac{\partial ^2 \chi^2}{\partial e \partial \omega} & \frac{\partial ^2 \chi^2}{\partial e^2} & \frac{\partial ^2 \chi^2}{\partial e \partial T_0} & \frac{\partial ^2 \chi^2}{\partial e \partial P} \\
    \frac{\partial ^2 \chi^2}{\partial T_0 \partial \gamma} & \frac{\partial ^2 \chi^2}{\partial T_0 \partial K} & \frac{\partial ^2 \chi^2}{\partial T_0 \partial \omega} & \frac{\partial ^2 \chi^2}{\partial T_0 \partial e} & \frac{\partial ^2 \chi^2}{\partial T_0^2} & \frac{\partial ^2 \chi^2}{\partial T_0 \partial P} \\
    \frac{\partial ^2 \chi^2}{\partial P \partial \gamma} & \frac{\partial ^2 \chi^2}{\partial P \partial K} & \frac{\partial ^2 \chi^2}{\partial P \partial \omega} & \frac{\partial ^2 \chi^2}{\partial P \partial e} & \frac{\partial ^2 \chi^2}{\partial P \partial T_0} & \frac{\partial ^2 \chi^2}{\partial P^2}
\end{bmatrix}\]
All second order partial derivatives are found numerically using central differences, from the following analytic first order derivatives
    \[\frac{\partial \chi^2}{\partial \gamma} = 2\sum_{i=1}^{N}{[V(t_i) - V_i]}\,w_i\]
    \[\frac{\partial \chi^2}{\partial K} = 2\sum_{i=1}^{N}{[V(t_i) - V_i][\cos(v + \omega) + e\cos\omega]}\,w_i\]
    \[\frac{\partial \chi^2}{\partial \omega} = -2K\sum_{i=1}^{N}{[V(t_i) - V_i][\sin(v+\omega)+e\sin\omega]}\,w_i\]
    \[\frac{\partial \chi^2}{\partial e} = 2K\sum_{i=1}^{N}{[V(t_i) - V_i][\cos(v+\omega)-\sin(v+\omega) \frac{\partial v}{\partial e}]}\,w_i\]
    \[\frac{\partial \chi^2}{\partial T_0} = -2K\sum_{i=1}^{N}{[V(t_i) - V_i]\sin(v+\omega)\frac{\partial v}{\partial T_0}}\,w_i\]
    \[\frac{\partial \chi^2}{\partial P} = -K\sum_{i=1}^{N}{[V(t_i) - V_i]\sin(v + \omega)\frac{\partial v}{\partial P}}\,w_i\]
Here $\frac{\partial v}{\partial e}$, $\frac{\partial v}{\partial T_0}$, and $\frac{\partial v}{\partial P}$ are given by
\begin{equation*}
\begin{split}
    \frac{\partial v}{\partial e} = 2\cos^2\Big(\frac{v}{2}\Big)\Bigg[\sqrt{\frac{1-e}{1+e}}\frac{1}{(1-e)^2}\tan\Big(\frac{E}{2}\Big) \\
    + \frac{1}{2}\sec^2\Big(\frac{E}{2}\Big)\sqrt{\frac{1+e}{1-e}}\frac{\sin{E}}{(1-e\cos{E})}\Bigg]
\end{split}
\end{equation*}
    \[\frac{\partial v}{\partial T_0} = \frac{1+\cos{v}}{1+\cos{E}}\sqrt{\frac{1+e}{1-e}}\frac{-n}{(1-e\cos{E})}\]
    \[\frac{\partial v}{\partial P} = -\frac{2\pi}{P^2}\frac{1+\cos{v}}{1+\cos{E}}\sqrt{\frac{1+e}{1-e}}\frac{t_i-T_0}{1-e\cos{E}}\]
    \\~\\


\end{document}